\documentclass[epj,referee]{svjour}
\usepackage{latexsym}
\usepackage{graphics}
\usepackage{amssymb}
\begin{document}
\title{Experimental study of the delayed threshold phenomenon in a Class-A VECSEL}
\author{A. El Amili\inst{1} \thanks{\emph{Present address:} Institut de Physique de Rennes, UMR CNRS-Universit\'e de Rennes I 6251, 35042 Rennes Cedex, France} \and G. Gredat\inst{1} \and M. Alouini\inst{2} \and I. Sagnes\inst{3} \and F. Bretenaker\inst{1}% etc
}                     % Do not remove
\institute{Laboratoire Aim\'e-Cotton, CNRS-Universit\'e Paris Sud 11, 91405 Orsay Cedex, France \and Institut de Physique de Rennes, UMR CNRS-Universit\'e de Rennes I 6251, 35042 Rennes Cedex, France \and Laboratoire de Photonique et Nanostructures, CNRS, 91460 Marcoussis, France}
\date{Received: date / Revised version: date}
\abstract{An experimental study of the delayed threshold phenomenon in a Vertical Extended Cavity Semiconductor Emitting Laser is carried out. Under modulation of the pump power, the laser intensity exhibits a hysteresis behavior in the vicinity of the threshold. The temporal width of this hysteresis is measured as a function of the modulation frequency, and is proved to follow the predicted scaling law. A model based on the rate equations is derived and used to analyze the experimental observations. A frequency variation of the laser around the delayed threshold and induced by the phase-amplitude coupling is predicted and estimated.
}
\maketitle
\section{Introduction}
\label{intro}
The laser threshold is a key concept which characterizes a given laser and marks the transition from vanishing intensity towards the lasing regime. Defining the threshold is a non trivial issue, which sparks off interesting discussions \cite{Bjork1994,Rice1994}. Classically, the threshold is defined as the operating point for which the unsaturated gain equals the losses. The laser threshold can then be considered as the bifurcation point between two distinct domains \cite{Erneux2010}. Below threshold, the zero intensity value is the stable solution while the non zero intensity value is the unstable solution. Above threshold, the stabilities of these two solutions are exchanged. From a more experimental point of view, the threshold can be considered as the region in which spontaneous and stimulated emissions have comparable weights. 
Furthermore, this bifurcation point is modified when a control parameter, such as, e.\ g., the intra-cavity losses or the gain, is varied in time across the threshold. The problem of the behavior of a laser under gain or loss modulation around threshold has been theoretically treated by Mandel and Erneux \cite{Mandel1984,Mandel1987,Erneux91}. When the laser is swept from below to above threshold, the laser equations predict a delayed threshold compared with the first threshold obtained by adiabatically sweeping the laser parameters. A linear sweeping across the bifurcation point also leads to a modification on the threshold profile itself. Indeed, the delayed threshold creates a sharp transition from zero intensity to laser oscillation.
To the best of our knowledge, this behavior was first noticed in the case of He-Ne lasers with swept cavity lengths by Mikhnenko \cite{Mikhnenko1971} and Aronowitz \cite{Aronowitz1972}. The first experimental investigations have been carried out by Sharpf \textit{et al.} \cite{Sharpf1987} with an Ar$^{+}$ laser in which the intra-cavity losses are linearly swept across the threshold.
A similar effect was studied in two different laser dynamics such as in the Class-A lasers \cite{Sharpf1987,chakmakjian89,vemuri91} and in the Class-B lasers \cite{arecchi89,ciofini90,balestri91,Tredicce2004}. 
When one of the control parameters is modulated, the laser intensity exhibits a hysteresis defined by two branches corresponding to the turning on and turning off the laser. In the vicinity of the threshold, a time lag between the two branches occurs. In the works cited above, it has been shown that the delay between the turn-on and turn-off branches depends on the modulation frequency according an inverse square root scaling law \cite{Mandel1987,Sharpf1987}.
 
In this article, we present an experimental study of the delayed threshold in a Class-A semiconductor laser. This work is a complementary investigation of the one reported in a reference \cite{Tredicce2004} on a Class-B semiconductor laser diode. Indeed, the analysis of the dynamical hysteresis, when the control of parameter is modulated, is quite difficult and the measurements are inaccurate in semiconductor lasers. This is due to the very short value of the photon lifetime in such lasers (due to the large intra-cavity losses and the smallness of the cavity) and also to the significant amplified spontaneous emission which may hide such an hysteresis cycle at low modulation frequency. Moreover, the spontaneous emission noise causes jitter on the time at which the laser turn-on. However, recently, Vertical External Cavity Semiconductor Lasers (VECSELs) have been shown to exhibit very long cold cavity decay times together with remarkably low noise levels \cite{Baili2007,Baili2008,laurain}. Moreover, these lasers can easily operate in the class-A dynamical regime \cite{Baili2009}. Thus, the aim of this paper is to report an experimental study of a VECSEL under modulation of the pump laser power. Indeed, such semiconductor lasers are well known for behaving quite similarly to the class-A gas lasers cited above. One can thus expect them to exhibit dynamical hysteresis around threshold. However, these lasers are semiconductor lasers in which the phase-amplitude coupling, described by Henry's $\alpha$ parameter \cite{henry82}, must play a significant role. This coupling must thus play a role in the frequency of the light field in the vicinity of the threshold where the laser does not reach its steady-state regime. To answer these questions, we present in the second section the analytical model describing the time evolution of the laser intensity close to threshold, and we propose a definition of the width of the dynamical hysteresis cycle. In the third section, we compare the  delays extracted from this model with our experimental measurements. Finally, we report an estimation of the frequency variation of the laser in the vicinity of the delayed threshold.
\section{Model}
\label{sec:1}
\begin{figure}
\resizebox{1.\columnwidth}{!}{%
  \includegraphics{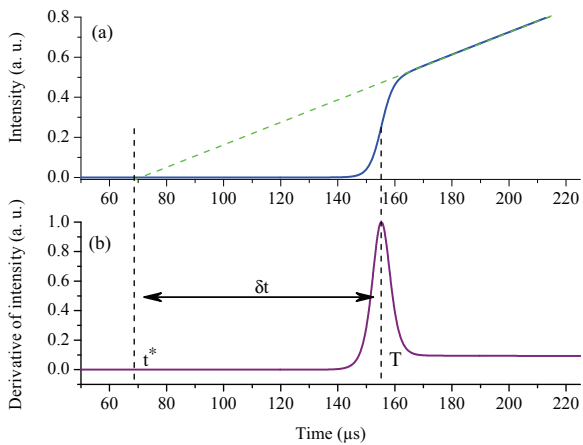}
}
\caption{(a) Full line: evolution of the laser intensity versus time, obtained from equation (\ref{eq_sol}) with $f_{\mathrm{m}}= 1$ kHz and $\eta= 146$ s$^{-1}$. Dashed line: corresponding ``adiabatic" laser intensity. It crosses the time axis at the date $t^{\ast}$. (b) First derivative of the intensity versus time. The maximum defines $T$. The delay $\delta t$ is defined as $T - t^{\ast}$.}
\label{fig:1}
\end{figure}

We consider here a single-frequency semiconductor laser in which the gain is produced by quantum wells. The gain is linearly swept around the laser threshold while the intra-cavity losses are kept constant. The laser is in the good cavity limit, also called class-A regime, meaning that the photon lifetime $\tau_{\mathrm{cav}}$ is much longer than the carriers lifetime $\tau_{\mathrm{c}}$. In this case, the time evolution of the carrier density can be adiabatically eliminated. The laser rate equations are then reduced to only one equation of evolution of the laser intensity:
\begin{equation}
\frac{dI}{dt} = \frac{I}{\tau_{\mathrm{cav}}}\left(\frac{r\left(t\right)}{1+\frac{I}{I_{\mathrm{sat}}}}-1\right)\ ,
\label{eq_laser}
\end{equation}
where $r\left(t\right)$ is the pumping parameter which is defined as the ratio between the unsaturated gain and the losses, and $I_{\mathrm{sat}}$ is the saturation intensity. The pumping parameter is varied linearly from below ($r<1$) to above ($r>1$) laser threshold. Since the pump laser power is linearly modulated, we express $r\left(t\right)$ as:
\begin{equation}
r\left(t\right) = r_{0} + \eta t\ .
\label{eq_tau_pompage}
\end{equation}
The sweep rate $\eta$ is simply given by $2 \left(r_{\mathrm{max}}-r_{0}\right) f_{\mathrm{m}}$, where $r_{0}$, $r_{\mathrm{max}}$, and $f_{\mathrm{m}}$ are respectively the initial pump parameter, the final pump parameter, and the modulation frequency. Close to threshold, $I\ll I_{\mathrm{sat}}$, and equation (\ref{eq_laser}) can be approximated by:
\begin{equation}
\frac{dI}{dt} \approx \frac{I}{\tau_{\mathrm{cav}}}\left(r\left(t\right)-\frac{I}{I_{\mathrm{sat}}}r\left(t\right)-1\right).
\label{eq_laserapprox}
\end{equation}
This last equation can be integrated to give the following solution \cite{Mandel1987}:
\begin{equation}
I\left(t\right) = \frac{e^{\frac{t}{\tau_{\mathrm{cav}}}\left(\frac{\eta}{2}t+r_{0}-1\right)}}{\frac{1}{I\left(0\right)}+\frac{r_{0}}{I_{\mathrm{sat}}\tau_{\mathrm{cav}}}\mathcal{D}\left(t\right)},
\label{eq_sol}
\end{equation}
where $\mathcal{D}\left(t\right)$ is given by:
\begin{equation}
\mathcal{D}\left(t\right) = \int_{0}^{t}e^{\frac{t'}{\tau_{\mathrm{cav}}}\left(\frac{\eta}{2}t'+r_{0}-1\right)}dt'.
\label{eq_integral}
\end{equation}
$\mathcal{D}\left(t\right)$ can be linked to the Dawson integral, leading to the following asymptotic expansion when $t\rightarrow \infty$:
\begin{equation}
\mathcal{D}\left(t\right) \simeq
\frac{e^{p t\left(p t + 2 q \right)}}{2 p\left(p t + q\right)}\left(1 + \frac{1}{2\left(p t+q\right)^{2}} + \frac{3}{4\left(p t + q\right)^{4}} + ...\right),
\label{eq_DL}
\end{equation}
where $p = \sqrt{\frac{\eta}{2\tau_{\mathrm{cav}}}}$ and $q = \sqrt{\frac{1}{2\eta\tau_{\mathrm{cav}}}}\left(r_{0}-1\right)$. 
The expression (\ref{eq_sol}) together with equation (\ref{eq_DL}) has been used to fit the experimental recordings of the evolution of the laser intensity. This model is also used to define the delay between the ``adiabatic" threshold, obtained with slow sweeping, and the delayed threshold. In this paper, we define two critical times as depicted in Figure\ \ref{fig:1}.
Figure \ref{fig:1}(a) reproduces the evolution of the laser intensity versus time in vicinity of the threshold. The full line is obtained using equation (\ref{eq_sol}): the laser intensity sharply passes from zero value to the non-zero value. Then, once above threshold, the laser intensity grows linearly with time.
The dashed line shows the ``adiabatic" evolution of the laser when the pumping parameter is slowly swept. This solution crosses the time axis at $t^{\ast}$ which corresponds to the time at which the laser reaches the ``adiabatic" threshold. At $t^{\ast}$ the losses are equal to the gain $\left[r\left(t^{\ast}\right)=1\right]$, leading to:
\begin{equation}
t^{\ast} = \frac{\left(1-r_{0}\right)}{\eta}.
\label{eq_tstar}
\end{equation}
One can define an another critical time $T$ corresponding to the time at which the first derivative of the intensity reaches his maximum as shown in Figure \ref{fig:1}(b). The difference between these two critical times characterizes the width of the dynamical hysteresis:
\begin{equation}
\delta t = T - t^{\ast}.
\label{eq_deltat}
\end{equation}

According to previous works \cite{Mandel1987,Sharpf1987,arecchi89,ciofini90,vemuri91}, the delay $\delta t$ depends on the modulation frequency, and it can be scaled as $\delta t \propto 1/ \sqrt{f_{m}}$. In the experimental part of the present paper, we compare this prediction with the experimental measurements performed with a VECSEL. This will also rule out other mechanisms leading to hysteresis cycles, such as those linked to polarization switching in VCSELs under current modulation, or thermal effect inducing hysteresis in the laser switch-on and switch-off points \cite{hong08,hong10}.
\section{Experiment}
\label{sec:2}
\subsection{Experimental set-up}
\begin{figure}
\resizebox{1.\columnwidth}{!}{%
  \includegraphics{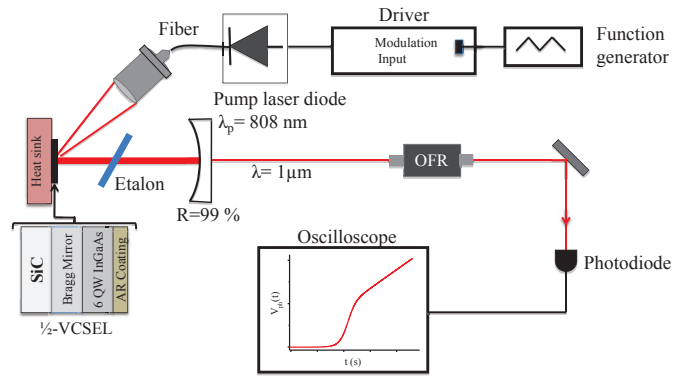} 
}
\caption{Sketch of the experimental set-up. The VECSEL cavity is in a planar-concave configuration. The laser is optically pumped with a pump laser diode. The current of this latter is modulated with a triangular function signal. The laser output is detected with a fast photodiode. The photodiode voltage is then recorded by an oscilloscope.}
\label{fig:2}
\end{figure}
The laser used in the present experiment is a single-frequency VECSEL which operates at $\sim$1 $\mu$m (see Fig. \ref{fig:2}). The planar-concave cavity is formed by a 1/2-VCSEL which provides the gain and acts as high reflectivity planar mirror, and a spherical output mirror (10 cm radius of curvature and 99$\%$ reflectivity). The cavity length is $L \lesssim 10\ \mathrm{cm}$. In the 1/2-VCSEL, the gain is produced by six InGaAs/GaAsP strained quantum wells (QWs) grown on a Bragg mirror which is bonded onto a SiC substrate in order to dissipate the heat towards a Peltier cooler. This multilayered stack is covered by an antireflection coating to prevent from any coupled cavity effect. The gain is broad ($\sim\,6$ THz bandwidth), spectrally flat, and has been  optimized to reach a low threshold \cite{laurain}. In order to force the VECSEL to oscillate on a single longitudinal mode, we insert an uncoated glass \'etalon (200 $\mu$m thick) inside the cavity. The laser is optically pumped at 808 nm. The current supply of the pump laser diode is modulated using a triangular voltage signal provided by a function generator. Starting below the VECSEL threshold $r_{0}<1$, the pumping rate increases linearly up to the maximum pumping parameter $r_{\mathrm{max}}$, before decreasing linearly back to its initial value. The optical signal is detected with a fast photodiode (rise-time $\sim$ 13 ns). The photodiode signal is then recorded with an oscilloscope.
\subsection{Dynamical hysteresis effect}
\begin{figure}
\resizebox{1.\columnwidth}{!}{%
  \includegraphics{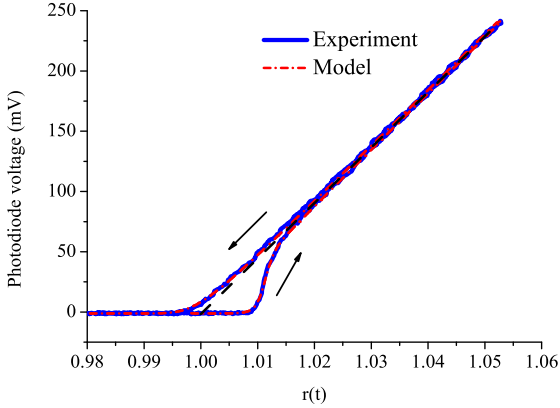}
}
\caption{Dots: measurements of the laser signal. The curves are the result of the averaging of five successive recordings. Full line: theoretical evolution of the laser signal, obtained from equation (\ref{eq_laser}) with $f_{\mathrm{m}}=$ 1 kHz, $\eta=$ 140 s$^{-1}$, and $\tau_{\mathrm{cav}}= 60\,\mathrm{ns}$. Dashed line: ``adiabatic" solution.}
\label{fig:3}
\end{figure}

The Figure\ \ref{fig:3} reproduces the evolution of the laser power versus pumping rate. The arrows indicate the direction of the evolution. The dots represent the experimental data recorded for a modulation frequency equal to 1 kHz. They clearly exhibit  the dynamical hysteresis evidenced by the existence of two different branches. The lower branch shows a sharp increase with a threshold delayed from the dashed line which represents the ``adiabatic" solution. This latter line crosses the horizontal axis at $r = 1$ (non-delayed threshold). The upper branch is the evolution of the laser intensity when the pumping rate decreases down to its initial value $r_{0}\sim 0.98$. By contrast with the lower branch, the upper one decreases smoothly towards zero.\\
The value of $\tau_{\mathrm{cav}}$ is obtained by fitting the lower branch of the experimental data using equations (\ref{eq_sol}) and (\ref{eq_DL}). Using the data of Figure\ \ref{fig:3}, we obtain $\tau_{\mathrm{cav}}\approx 60$ ns. This value is consistent with the length and losses of the cavity. The full line in Figure\ \ref{fig:3} is obtained by numerical integration of Equation\ (\ref{eq_laser}) with this value. This leads to a very good agreement with the experimental data.

The behavior of the laser intensity in Figure\ \ref{fig:3} can be physically understood in the following way. For the lower branch (increasing pumping parameter), the response time of the laser, which diverges close to threshold due to the critical slowing down phenomenon, restrains the laser from responding fast enough to the fact that it is suddenly above threshold, leading to a delay in its actual threshold. On the contrary, for the upper branch (decreasing pumping parameter), the gain suddenly becomes smaller than the losses while the cavity is filled with light. Thus, the intracavity intensity decreases exponentially with a decay time governed by the cavity photon lifetime.
\subsection{Width of the dynamical hysteresis cycle}
In this section, we present our measurements of the evolution of the width $\delta t$ of the hysteresis cycle when the pump power modulation parameters are varied. We remind that we define $\delta t$ as in Figure\ \ref{fig:1}. According to Mandel \cite{Mandel1987}, Arecchi \textit{et al.} \cite{arecchi89}, and our numerical simulations of equation (\ref{eq_laser}), we expect $\delta t$ to scale as the inverse square root of $f_{\mathrm{m}}$.

\begin{figure}
\resizebox{1.\columnwidth}{!}{%
  \includegraphics{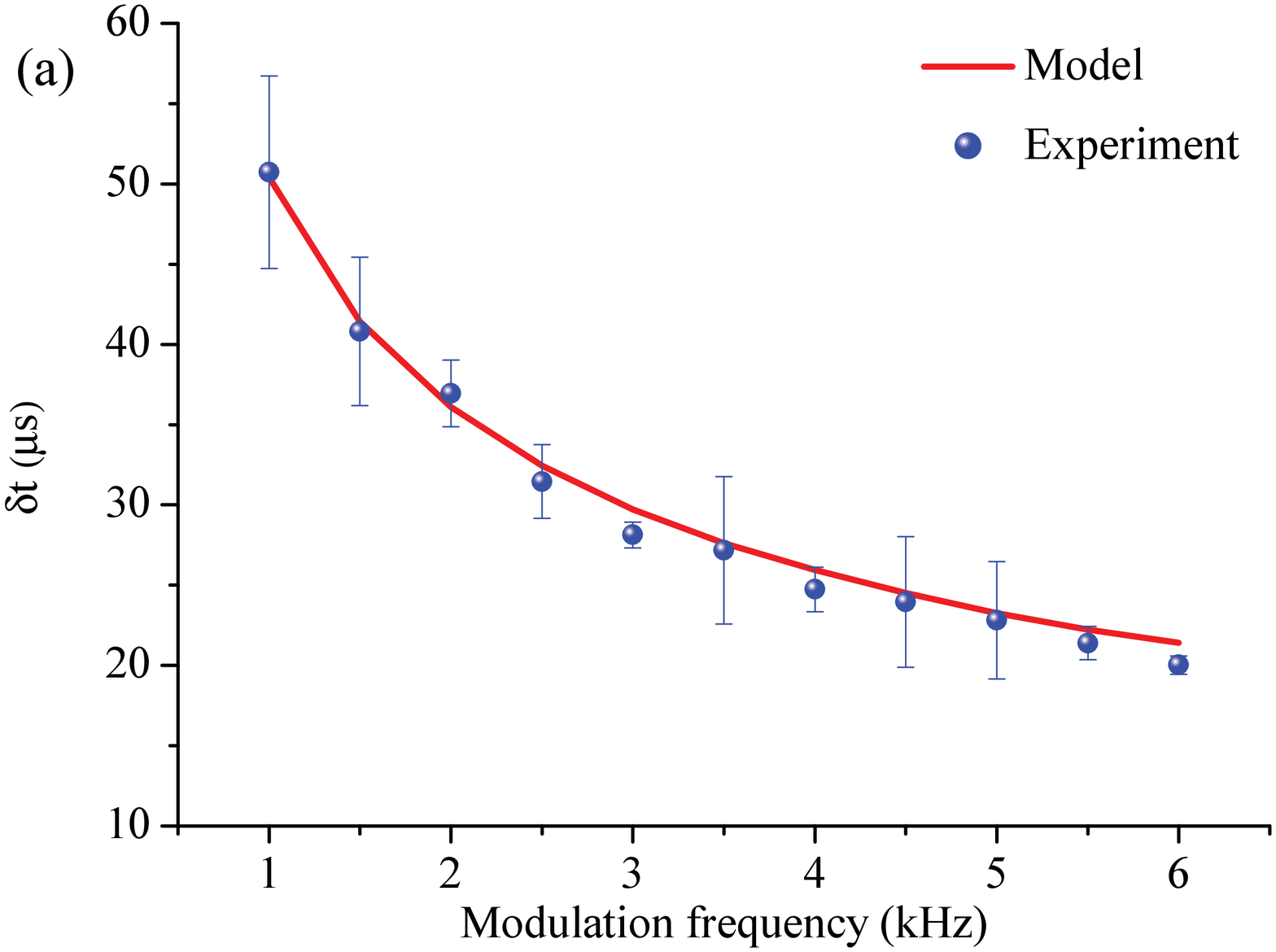}
}
\resizebox{1.\columnwidth}{!}{%
  \includegraphics{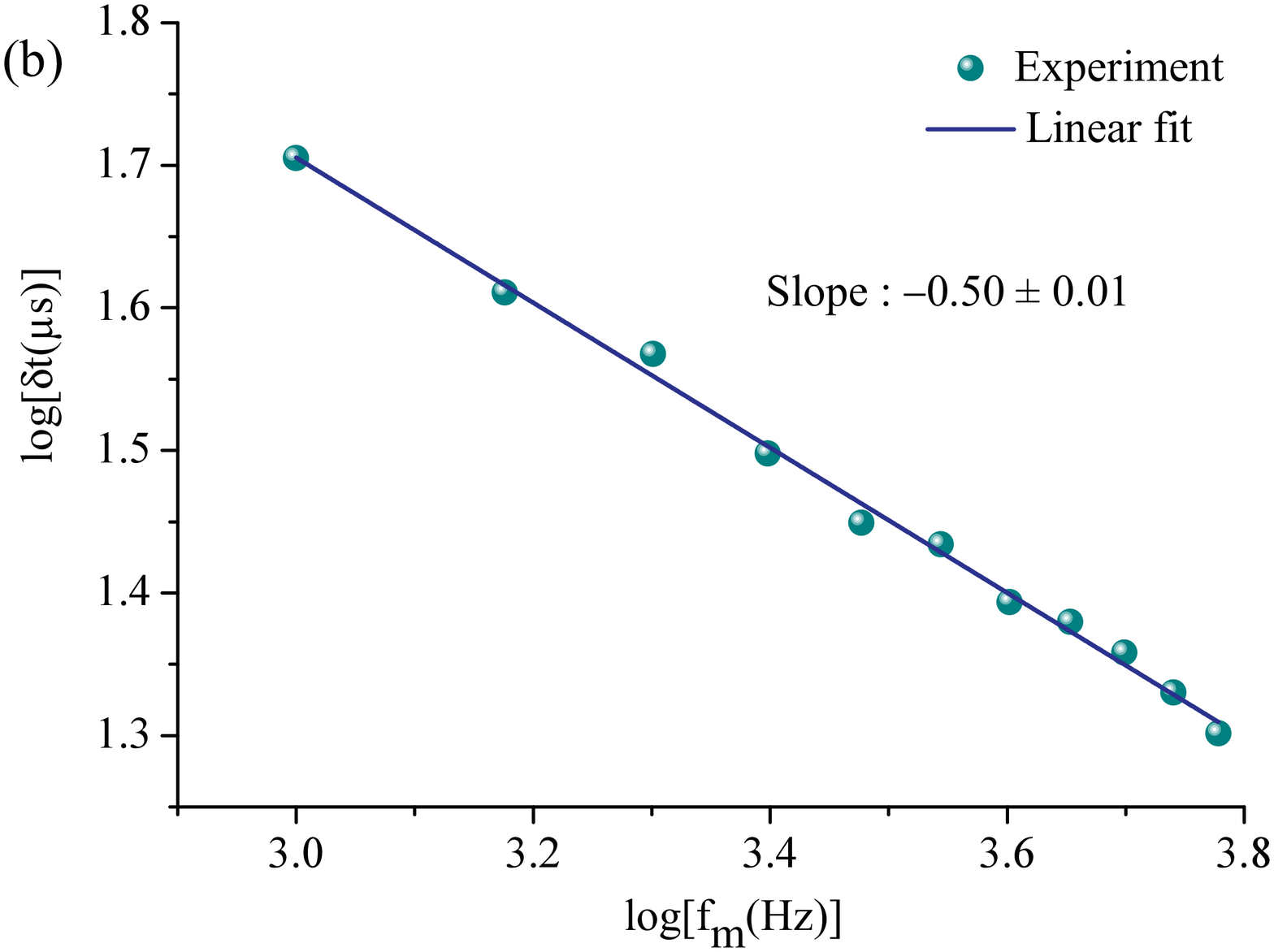}
}
\caption{(a) Dots: measured width $\delta t$ of the hysteresis cycle as a function the modulation frequency $f_{\mathrm{m}}$. Full line: numerical simulations. (b) Same data on a log-log plot. Full line: linear fit leading to a slope equal to $-0.50 \pm 0.01$.}
\label{fig:2_3}
\end{figure}

To check this behavior, the laser signal evolution is recorded for several values of the modulation frequency ranging from 1 kHz to 6 kHz by steps of 500 Hz. For each modulation frequency, we record a set of five traces of the laser signal, which are averaged. We extract $\delta t$ from this average. Figure\ \ref{fig:2_3}(a) reproduces the evolution of $\delta t$ versus $f_{\mathrm{m}}$. The experimental data are represented by the dots and show that the delay decreases when the modulation frequency increases. This behavior is consistent with the results reported previously for gas lasers \cite{Mandel1987,arecchi89,ciofini90}. The full line in Figure \ref{fig:2_3}(a) corresponds to the theoretical values of $\delta t$ obtained by simulating equation (\ref{eq_laser}) with $\tau_{\mathrm{cav}}= 60\,\mathrm{ns}$. We obtain a very good agreement with the experimental measurements. Figure \ref{fig:2_3}(b) shows the same data on a log-log plot. A linear fit leads to a negative slope of  $-0.50 \pm 0.01$.
This is in excellent agreement with the expected $\delta t \sim 1 /\sqrt{f_{\mathrm{m}}}$ scaling law. To our knowledge, theses results are the first such measurements carried out with a Class-A semiconductor laser.
\subsection{Estimation of the laser frequency chirp at threshold}
\begin{figure}
\resizebox{1.\columnwidth}{!}{%
  \includegraphics{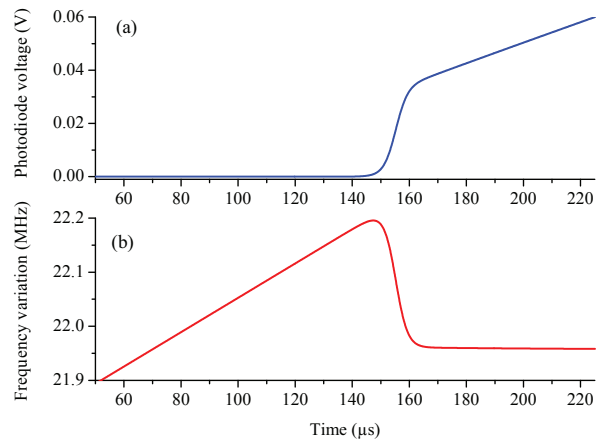}
}
\caption{(a) Time evolution of the laser signal at the modulation frequency $f_{\mathrm{m}}=$ 1 kHz. (b) Frequency variation as function the time calculated with Henry's parameter $\alpha =$ 7.}
\label{fig:4}
\end{figure}

Up to now, we have seen that our class-A semiconductor laser behaves exactly like the gas lasers in which dynamical hysteresis has been observed up to now. However, one salient feature of semiconductor lasers is the existence of a non negligible phase-amplitude coupling effect summarized in Henry's factor $\alpha$. Since around the delayed threshold, the population inversion has a non-linear variation as a function of time, we can thus expect the laser phase to be affected by this effect. The subsequent frequency variation can be estimated by rewriting equation (\ref{eq_laser}) for the slowly varying complex amplitude $\mathcal{A}$ of the laser field:
\begin{equation}
\frac{d\mathcal{A}}{dt}  =  \frac{\mathcal{A}}{2\tau_{\mathrm{cav}}}\left(\left(1-i\alpha\right)\frac{r(t)}{1+\frac{I}{I_{\mathrm{sat}}}}-1\right) \ ,\label{eqn-A}
\end{equation}
where we have used the fact that our laser is a Class-A laser, where we have chosen the $e^{-i\omega t}$ time dependence for the complex field, and where we suppose for simplicity that $I= \left|\mathcal{A}\right|^2$. The laser frequency deviation $\delta\nu$ with respect to the cold cavity eigenfrequency is obtained by taking the imaginary part of equation (\ref{eqn-A}), leading to:
\begin{equation}
\delta\nu\left(t\right) = \frac{\alpha}{4\pi\tau_{\mathrm{cav}}}\frac{r\left(t\right)}{1+\frac{I}{I_{\mathrm{sat}}}}.
\label{eq_deltanu}
\end{equation}
When the laser has reached steady state, one has $\frac{r\left(t\right)}{1+\frac{I}{I_{\mathrm{sat}}}}= 1$ leading to the following frequency deviation with respect to the cold cavity frequency:
\begin{equation}
\delta\nu_{\mathrm{st}} = \frac{\alpha}{4\pi\tau_{\mathrm{cav}}}.
\label{eq_deltast}
\end{equation}
Figure\ \ref{fig:4} reproduces the evolutions of the intensity and the frequency deviation versus time according to equations. (\ref{eq_laser}) and (\ref{eq_deltanu}), with the same parameters as in Figure \ref{fig:1}.
We have also taken $\alpha= 7$. The frequency deviation follows the time evolution of the population inversion. It reaches its maximum value during the hysteresis cycle, i. e., when the population inversion is different from its threshold value, i. e., when the laser intensity is very different from its adiabatic value. Later, when the laser intensity is back to its adiabatic value, the frequency variation reaches its steady state value $\delta\nu_{\mathrm{st}}\approx$ 21.95 MHz. Then, the population inversion remains equal to the population inversion at threshold. Notice that the expected peak-to-peak variation of the frequency during the delayed threshold phenomenon is of the order of 250~kHz, making it quite difficult to observe, particularly because it occurs on short time scales (about $10\;\mu$s) and for very low output powers.
\section{Conclusion}
In this paper we have reported an experimental observation of the delayed threshold phenomenon in a Class-A semiconductor laser. We have shown that as long as the delayed threshold phenomenon is concerned, our VECSEL behaves just like CO$_{2}$ or Ar$^{+}$ laser systems in which this effect had already been studied. For example, we have shown experimentally and theoretically that the time lag $\delta t$ scales as $1/\sqrt{f_{\mathrm{m}}}$, where $f_{\mathrm{m}}$ is the modulation frequency of the pump. Our experimental results are in very good agreement with a model based on the rate equations for a Class-A laser. We have also predicted the existence of a frequency variation which occurs in the vicinity of the delayed threshold just while the laser starts lasing. This frequency variation exists only in semiconductor lasers due to the fact $\alpha\neq 0$. This phenomenon deserves to be observed with a dedicated experimental apparatus. Indeed, according to the simple dependence of $\delta\nu$ versus time with respect to $\alpha$, such an apparatus would offer a direct and accurate way to measure the Henry factor in VCSEL active media.
% BibTeX users please use
% \bibliographystyle{}
% \bibliography{}

\begin{thebibliography}{}
%
\bibitem{Bjork1994}
G. Bj$\ddot{o}$rk, A. Karlsson, Y. Yamamoto, Phys. Rev. A \textbf{50}, 1675 (1994)
%
\bibitem{Rice1994}
P.R. Rice, H.J. Carmichael, Phys. Rev. A \textbf{50}, 4318 (1994)
%
\bibitem{Erneux2010}
T. Erneux, P. Glorieux, \textit{Laser Dynamics} (Cambridge University Press, Cambridge 2010)
%
\bibitem{Mandel1984}
P. Mandel, T. Erneux, Phys. Rev. Lett. \textbf{53}, 1818 (1984)
%
\bibitem{Mandel1987}
P. Mandel, Optics Commm. \textbf{64}, 549 (1987)
%
\bibitem{Erneux91}
T. Erneux, P. Mandel, Optics Commm. \textbf{85}, 43 (1991)
%
\bibitem{Mikhnenko1971} 
G.A. Mikhnenko, E.D. Protsenko, E.A. Sedoi, M.P. Sorokin, Opt. Spectrosc. \textbf{30}, 65 (1970)
%
\bibitem{Aronowitz1972}
F. Aronowitz, Appl. Opt. \textbf{11}, 405 (1972)
%
\bibitem{Sharpf1987}
W. Sharpf, M. Squicciarini, D. Bromley, C. Green, J.R. Tredicce, L.M. Narducci, Optics Commm. \textbf{63}, 344 (1987)
%
\bibitem{arecchi89}
F.T.~Arecchi, W. Gadomski, R. Meucci, J.A. Roversi, Optics Commm. \textbf{70}, 155 (1989)
%
\bibitem{ciofini90}
M. Ciofini, R. Meucci, F.T. Arecchi, Phys. Rev. A \textbf{42}, 482 (1990)
%
\bibitem{balestri91}
S. Balestri, M. Ciofini, R. Meucci, F.T. Arecchi, P. Colet, M. San Miguel, S. Balle, Phys. Rev. A \textbf{44}, 5894 (1991)
%
\bibitem{chakmakjian89}
S.H. Chakmakjian, S. Papademetriou, K. Koch, C.R. Stroud Jr., Phys. Rev. A \textbf{40}, 1858 (1989)
%
\bibitem{vemuri91}
G. Vemuri, Phys. Rev. A \textbf{85}, 36 (1991)
%
\bibitem{Tredicce2004}
J.R. Tredicce, G.L. Lippi, P. Mandel, B. Charasse, A. Chevalier, B. Picqué, Am. J. Phys. \textbf{72}, 799 (2004)
%
\bibitem{Baili2007}
G. Baili, M. Alouini, C. Moronvalle, D. Dolfi, F. Bretenaker, I. Sagnes, A. Garnache, Opt. Lett. \textbf{32}, 650 (2007)
%
\bibitem{Baili2008}
G. Baili, F. Bretenaker , M. Alouini, L. Morvan, D. Dolfi, I. Sagnes, J. Lightwave Technol. \textbf{26}, 952 (2008)
%
\bibitem{laurain}
A. Laurain, M. Myara, G. Beaudoin, I. Sagnes, A. Garnache, Opt. Express \textbf{17}, 9503 (2009)
%
\bibitem{Baili2009}
G. Baili, M. Alouini, T. Malherbe, D. Dolfi, I. Sagnes, F. Bretenaker, Europhys. Lett. \textbf{87}, 4405 (2009)
%
\bibitem{henry82}
C. Henry, IEEE J. Quantum Electron. \textbf{18}, 259 (1982)
%
\bibitem{hong08}
Y. Hong, J. Paul, P.S. Paul, K.A. Shore, IEEE J. Quantum Electron. \textbf{44}, 30 (2008)
%
%
\bibitem{hong10}
Y. Hong, C. Masoller, M.S. Torre, S. Priyadarshi J. Paul, A.A. Qader, P.S. Paul, K.A. Shore, Opt. Lett. \textbf{35}, 3688 (2010)
%Author, \textit{Book title} (Publisher, place year) page numbers
\end{thebibliography}

\end{document}